# Imaging gate-tunable single-layer molecular kagome lattices

Ziyi Wang[1,2,3,4]†, Rafal Zuzak[1]†, Weichen Tang[1,3]†, Haowen Pang[1], Yutang Qiao[1,7], Dhanvanth Balakrishnan[1], Yuqiao Yan[1,8], Aining Hu[1], Andrew Sun[1], Takashi Taniguchi[5], Kenji Watanabe[6], Feng Wang[1,3,4]*, Steven G. Louie[1,3]*, Michael F. Crommie[1,3,4]*

**Affiliations:**

[1]Department of Physics, University of California, Berkeley, CA 94720, USA.

[2]Department of Chemistry, Columbia University, New York, NY 10027, USA.

[3]Materials Sciences Division, Lawrence Berkeley National Laboratory, Berkeley, CA 94720, USA.

[4]Kavli Energy NanoSciences Institute at the University of California Berkeley and the Lawrence Berkeley National Laboratory, Berkeley, CA 94720, USA.

[5]Research Center for Materials Nanoarchitectonics, National Institute for Materials Science, Tsukuba, Japan.

[6]Research Center for Electronic and Optical Materials, National Institute for Materials Science, Tsukuba, Japan.

[7]Present address: Department of Physics, Yale University, New Haven, CT 06511, USA.

[8]Present address: Joseph Henry Laboratories and Department of Physics, Princeton University, Princeton, NJ 08544, USA.

*Emails: crommie@berkeley.edu, sglouie@berkeley.edu, fengwang76@berkeley.edu

†These authors contributed equally to this work.

**Abstract:**

Kagome lattices provide a simple geometry for engineering electronic states that exhibit novel quantum behavior. In conventional kagome materials, however, the relevant states are often embedded in chemically complex three-dimensional crystals and are difficult to isolate or tune. Molecule-based kagome networks offer structural precision and tunability, but have largely been confined to metallic substrates incompatible with 2D devices. Here we report growth of single-layer $Ni_3(HITP)_2$ metal–organic frameworks (MOFs) on van der Waals substrates including graphene-based field-effect devices using metal-salt molecular beam epitaxy. Low-temperature scanning tunneling spectroscopy resolves kagome-derived flat-band and van Hove singularity-based spectral features consistent with first-principles calculations. Gating hybrid 2D MOF devices allows us to energetically shift kagome spectral features, thus establishing electrostatic control of chemically programmable 2D kagome electronic lattices.

## Main

2D kagome lattices provide a route to achieve flat electronic bands, Dirac crossings, van Hove singularities, and the novel correlated and topological physics associated with these band structure features.[1-3] A central challenge, however, is the realization of 2D kagome systems that are structurally well-defined and have a continuously tunable Fermi energy ($E_F$), since this is not yet achievable using conventional solid-state kagome materials.[3-6] Bulk kagome metals such as $AV_3Sb_5$ and $RMn_6Sn_6$ exhibit novel superconductivity, charge order, and topological magnetism,[7-10] but their electronic structures typically deviate significantly from ideal 2D kagome characteristics and they are not easily integrated with 2D gate-tunable device platforms.[3,4,11,12] Molecular kagome networks offer an alternate route for accessing kagome-based physics because their lattice geometry and orbital structure can be flexibly programmed through chemical synthesis. Metal-organic frameworks (MOFs), for example, allow atomically-precise 2D kagome lattices to be engineered from organic linkers and metal coordination nodes.[13-18] Up to now, however, kagome MOFs have mainly been grown on bulk metal substrates that are incompatible with device-based electronic tunability.[19-22]

We have overcome this limitation by growing single-layer kagome MOFs directly on semiconducting and semimetal bulk van der Waals (vdW) substrates, as well as on monolayer (ML) graphene/hexagonal boron nitride (hBN) field-effect (FET) devices. Combined scanning tunneling microscopy (STM), scanning tunneling spectroscopy (STS), and first principles electronic structure calculations show that our bottom-up fabricated MOFs exhibit expected 2D kagome electronic structure features. Our MOF growth was accomplished using a novel metal-salt molecular beam epitaxy (metal-salt MBE) approach to obtain well-ordered covalently-bonded kagome domains at the surface of graphite and $WSe_2$ bulk crystals as well as ML-graphene/hBN devices. Here the transition metal component of our MOF is deposited using low-temperature metal bromide evaporation rather than more conventional high-temperature elemental e-beam evaporation.[20,23] STM spectroscopy of single-layer kagome MOFs grown at the surfaces of graphite and ML-graphene/hBN show signatures of all the major features expected of a kagome system, including flat bands, van Hove singularities, and Dirac points. These features compare well to *ab initio* DFT simulations of the electronic structure of isolated single-layer kagome MOFs having identical chemical structure. MOFs grown on ML-graphene/hBN devices have an energy gap of ~ 0.7 eV and exhibit a valence flat band that appears only 0.2 eV below the graphene Dirac point. Electrostatic gating of these devices causes the kagome band structure features to rigidly shift in energy by ±0.12 eV, comparable to the energy shift of bare graphene FET electronic features under identical gating conditions.[24,25] These results establish a new device-integrated, chemically programmable 2D molecular kagome platform in which local electronic structure can be imaged and electrostatically tuned.

## Design of a single-layer molecular kagome lattice

To realize a chemically-defined kagome lattice compatible with on-surface synthesis and STM imaging in 2D devices, we selected the previously reported MOF $Ni_3(HITP)_2$ (HITP = 2,3,6,7,10,11-hexaiminotriphenylene).[26] Here square-planar $NiN_4$ coordination nodes are linked by extended π-conjugated HITP ligands to form a 2D kagome network of corner-sharing triangles (Fig. 1A). Density functional theory (DFT) simulations of a freestanding $Ni_3(HITP)_2$ monolayer (Fig. 1B) reveal a three-band conduction manifold that captures the hallmarks of a kagome lattice: a flat band (FB1) with nearby Dirac point (DP) and flanking van Hove singularities (VHS1 and VHS2). Wannier analysis maps this conduction manifold onto three symmetry-related orbitals centered on the three $NiN_4$ kagome sites that are connected by nearest neighbor and next nearest-neighbor hopping (Supplementary Materials Fig. S1). Valence band states are separated from the conduction bands by an energy gap and show a slightly distorted kagome manifold. Here a flat band (FB2) lies below the charge neutrality point and above a van Hove singularity (VHS3). The more complex valence-band dispersion is consistent with stronger mixing between Ni *d* orbitals and HITP-derived states. At even lower energy a second valence manifold can be seen that exhibits a third flat band (FB3).

## Metal-salt molecular beam epitaxy of $Ni_3(HITP)_2$ on van der Waals substrates

We first attempted to grow $Ni_3(HITP)_2$ on highly oriented pyrolytic graphite (HOPG) using the conventional method of co-depositing elemental Ni and 2,3,6,7,10,11-hexaaminotriphenylene (HATP) molecules (see Supplementary Materials and Methods).[20,23] However, this resulted in problematical Ni clustering and disordered molecular adsorption, with no extended kagome lattice formation. We then turned to a new metal-salt MBE growth scheme in which anhydrous $NiBr_2$ and HATP precursors are supplied from separate source crucibles and reacted directly on the growth substrate which is held at elevated temperature ($T \approx 250$ °C) to form $Ni_3(HITP)_2$ (Fig. 1C). This was motivated by the idea that $NiBr_2$ might suppress metallic Ni aggregation by delivering Ni in a halide-coordinated ionic form that can react with HATP during MOF nucleation.

This new growth scheme resulted in well-ordered single-layer $Ni_3(HITP)_2$ kagome lattice domains on HOPG separated by grain boundaries (Fig. 2A, white dashed lines). Higher resolution STM topographs (Fig. 2B) show the kagome lattice constant to be ~19 Å, consistent with the structural model of Fig. 1A. We identify two common defect types: excess HITP molecules residing in lattice pores (red dashed circle) and HITP vacancies (yellow dashed circle), with an overall defect density of $\sim 10^{12}$ $cm^{-2}$. Fourier transforms of our STM images yield sixfold-symmetric Bragg peaks (Supplementary Materials Fig. S2), consistent with long-range kagome order.[27,28]

Single-layer kagome lattice growth was also achieved for bulk *p*-doped $WSe_2$ substrates (Fig. 2, C and D), demonstrating that metal-salt MBE is compatible with different vdW materials. The growth morphology differs, however, depending on the substrate. On HOPG $Ni_3(HITP)_2$ forms large islands that merge (Fig. 2A), whereas on $WSe_2$ $Ni_3(HITP)_2$ grows in the form of numerous smaller islands that have dendritic character (Fig. 2C). This difference is consistent with the existence of a higher density of growth nucleation sites on $WSe_2$ due to dopants and surface defects that do not occur for graphite (Supplementary Materials Fig. S4).

**Local spectroscopic resolution of kagome-derived electronic states**

We next tested whether predicted kagome electronic structure can be resolved directly by local STS of $Ni_3(HITP)_2$ grown on HOPG. Fig. 3A shows a representative d$I$/d$V$ spectrum acquired in the interior of a $Ni_3(HITP)_2$ domain (STS spectra were acquired at least 10 nm from edges to avoid possible edge effects). The spectrum exhibits pronounced features for filled and empty states that are separated by an energy gap of $\Delta E \approx 0.7$ eV. On the conduction band side, a prominent peak is seen at $V_s \approx +0.9$ V (*feature* 1) that lies above an "M-shaped" structure with two shoulders (*features* 2 and 4) flanking a central dip (*feature* 3). On the valence band side the topmost feature is a strong peak at $V_s \approx -0.2$ V (*feature* 5), below which several smaller resonances/shoulders lie within a few tens of mV. We label the most prominent of these *feature* 6. As $V_s$ is reduced the d$I$/d$V$ signal dips to a minimum at $V_s \approx -0.5$ V but then rises to form a prominent resonance centered at $V_s \approx -0.8$ V (*feature* 7).

We spatially resolved the MOF kagome states by recording constant-height d$I$/d$V$ maps at the energies of the most prominent spectral peaks (i.e., *features* 1 and 5). As shown in Fig. 3B, the d$I$/d$V$ map of *feature* 1 ($V_s = +0.9$ V) in the conduction band is distributed broadly across the entire MOF framework, including substantial weight on the triphenylene ligands. In contrast, the state corresponding to *feature* 5 ($V_s = -0.2$ V) in the valence band is concentrated primarily on the $NiN_4$ nodes with very weak intensity on the organic linkers (Fig. 3C).

To understand the origin of these experimental STS features we calculated the total density of states (DOS) of a freestanding $Ni_3(HITP)_2$ monolayer (Fig. 3F). The simulated kagome DOS reproduces the main features of our data quite nicely. In the conduction band, for example, a flat-band peak (FB1) is seen that corresponds to experimental *feature* 1, while a pair of smaller peaks (VHS1 and VHS2) arising from van Hove singularities correspond to experimental *features* 2 and 4, and a theoretical Dirac point dip (DP) corresponds to experimental *feature* 3. The overall energy width of the theoretical conduction band features (~0.4 eV) reasonably matches the experimental width (~0.5 eV). On the valence band side the theoretical flat-band peak marked FB2 corresponds well to experimental *feature* 5. Just below FB2 the match is not as good, but there is a rough correspondence between the theoretical van Hove singularity (VHS3) and experimental *feature* 6. Below VHS3 both theory and experiment show a significant reduction in spectral density before rising in a strong resonance that shows agreement between theoretical FB3 and experimental *feature* 7. The general good agreement between theory and experiment extends to the spatial distribution of the kagome electronic states, as seen by comparison of the theoretical LDOS for flat band FB1 (Fig. 3D) and flat band FB2 (Fig. 3E) to the experimental d$I$/d$V$ images of Fig. 3B and C. Here the filled-state LDOS concentrates primarily on the $NiN_4$ centers whereas the empty-state LDOS extends across the organic ligand for both theory and experiment, including a prominent node at the ligand center.

While there is good agreement between theory and experiment for the main spectroscopic features of $Ni_3(HITP)_2$, there remain some discrepancies. For example, the experimental spectral density “tail” below *feature* 4 is much shorter than the tail seen below the simulated peak of VHS2. This highlights the difficulty in some STM experiments of determining precise band edges when bands exhibit energetically long tails, since the STM tip must be brought very close to the surface to capture small, evanescent features.[29,30] In these cases it is difficult to determine the “true” bandgap using STS, and so "functional" energy gaps are often defined instead.[31] For the kagome system studied here we define the functional energy gap as the separation between the highest points of the LDOS peaks nearest the valence and conduction band edges, since these features are robust and experimentally unambiguous. Following this protocol allows us to define an experimental energy gap between *features* 4 and 5 of 0.7 ± 0.1 eV, which we can then compare to the theoretical energy gap between peaks VHS2 and FB2 of 0.45 eV. The smaller calculated gap may partly reflect the well-known tendency of DFT–PBE to underestimate semiconductor bandgaps.[32-36] Although these peak separations also depend on the dispersions near the band edges, the overall correspondence between experiment and theory supports this interpretation.

**Device integration and electrostatic tuning of kagome-derived states**

Our new ability to grow high-quality MOFs on HOPG and semiconducting TMD surfaces opens the possibility of growing 2D kagome lattices directly on single-layer FET devices made from these substrate materials. Our motivation for doing this is that the gate-tunability of such devices provides a possible strategy for tuning the MOF Fermi energy ($E_F$) into interesting regions of the kagome band structure without having to introduce impurity dopants. In order to test the feasibility of this strategy we grew single-layer kagome MOFs on ML-graphene/hBN FET device substrates. The graphene device stacks (Fig. 4A) were assembled using a PVC-based dry pickup technique where graphene and hBN flakes are sequentially stacked and then transferred onto a $SiO_2$/Si substrate, with Si serving as a global back-gate. Cr/Au electrical contacts were then deposited by shadow-mask evaporation to reduce surface contamination (Fig. 4B), and then wire-bonded. Graphene devices were AFM-scratched for cleanliness[37,38] before putting them into the UHV chamber and annealing at 400 °C for 12 hours for additional cleanliness. Device surfaces were checked via STM in UHV to ensure atomic cleanliness (Fig. 4C) before initiating MOF growth in a connected chamber.

Single-layer $Ni_3(HITP)_2$ MOFs were grown directly on ML-graphene/hBN devices using the same $NiBr_2$/HATP metal-salt MBE strategy used for HOPG substrates, but with a higher substrate temperature of 320 °C compared to 250 °C for HOPG. The need for higher temperature here for good growth possibly reflects the lower thermal conductivity of the device stack relative to bulk HOPG or $WSe_2$. STM imaging after performing this growth protocol revealed well-ordered $Ni_3(HITP)_2$ kagome domains on the graphene device that are structurally indistinguishable from those obtained on HOPG except for the presence of a large-scale background graphene/hBN moiré pattern (Fig. 4D).

To assess substrate effects in the device geometry we calculated the electronic structure of $Ni_3(HITP)_2$ supported on monolayer graphene (Supplementary Materials Fig. S2). The MOF-projected bands retain the characteristic kagome manifolds of the freestanding monolayer, indicating weak hybridization with graphene over the relevant energy range. The Fermi level remains within the MOF-projected gap, with no evidence of appreciable graphene-induced carrier doping in the modeled neutral heterostructure. These results support using the freestanding MOF monolayer calculation to interpret the principal features of MOF/graphene device STS.

We used the Si back-gate of our $Ni_3(HITP)_2$/graphene/hBN device to electrostatically tune the electronic structure of a single-layer bottom-up-fabricated kagome lattice. Fig. 5A shows a cartoon sketch of how gating is expected to alter the electronic structure of $Ni_3(HITP)_2$/graphene assuming a simple "rigid band" model where increased gate-voltage ($V_G$) causes an increase in $E_F$. Fig. 5B shows a density map of the STM $dI/dV$ signal measured as a function of $V_s$ and $V_G$ with the STM tip held at a single point above the kagome overlayer. The vertical blue band bracketing $V_s = 0$ shows the same experimental kagome energy gap seen in Fig. 3A. To the left of this is a vertical red feature that arises from the high DOS of kagome *feature* 5 in the filled states (i.e., flat band 2 (FB2)). This feature has a negative slope with $V_s$, indicating that FB2 shifts *down* in energy as $V_G$ is *increased* (consistent with rigid band motion as sketched in Fig. 5A). Other kagome electronic structure features, such as *feature* 1 in the empty states (FB1) and *feature* 7 at lower $V_s$ (FB3) similarly shift down in energy with increased $V_G$ at a rate of $\partial E/\partial V_G \approx -3$ meV/V, thus demonstrating the overall gate-tunability of single-layer kagome electronic structure (See Supplementary Materials Fig. S5 for detailed FB peak shift analysis).

Closer inspection of Fig. 5B reveals a set of sharp resonances that move with *positive* slope as $V_s$ is increased (green arrows). We attribute these features to charging events involving localized states, most likely associated with defects that trap charge in or near the $Ni_3(HITP)_2$ layer.[39,40] These resonances are different from the "normal" DOS features discussed above and behave similar to quantum dot charging behavior observed previously that is characterized by "Coulomb diamonds" as $V_s$ and $V_G$ are varied.[41,42] In some cases we observe these charging features to be accompanied by abrupt local shifts of nearby kagome-derived spectral features (yellow arrows), consistent with local charge-induced modification of the electrostatic environment near the STM tip.

## Conclusion and outlook

We have demonstrated a new strategy for achieving single-layer MOF kagome lattices whose local electronic structure can be imaged and electrostatically tuned in a van der Waals device geometry. Metal-salt MBE enables MOF growth on weakly interacting substrates, avoiding the bulk metal substrates that have historically limited device control of on-surface-synthesized molecular networks. Low-temperature STS resolves kagome-derived spectral features, including narrow flat-band peaks and van Hove singularity features, consistent with first-principles calculations. Integration of metal-salt MBE growth of MOFs with ML-graphene/hBN FET devices enables observation of gate-dependent energy shifts of the kagome spectral density. Future improved devices should allow stronger gating that enables the device $E_F$ to be shifted directly into interesting regions of the 2D kagome band structure, such as flat bands, Dirac points, and van Hove singularities.

**Acknowledgement**

This work was primarily funded by the US Department of Energy (DOE), Office of Science, Basic Energy Sciences (BES), Materials Sciences and Engineering Division, under contract DE-AC02-05-CH11231 (Nanomachine program KC1203). This research used computational resources provided by the National Energy Research Scientific Computing Center (NERSC), a Department of Energy User Facility using NERSC award No. BES-ERCAP0037553, the Frontera computing project at the Texas Advanced Computing Center (TACC) at The University of Texas at Austin, which is supported by National Science Foundation award No. OAC-1818253. Z.W. acknowledges support from a Kavli ENSI Graduate Student Fellowship. R.Z. acknowledges fellowship support from the National Science Center, Poland, grant number 2024/55/D/ST3/01489. K.W. and T.T. acknowledge support from the JSPS KAKENHI (Grant Numbers 21H05233 and 23H02052), the CREST (JPMJCR24A5), JST and World Premier International Research Center Initiative (WPI), MEXT, Japan (for hBN crystal development and synthesis). We thank Dr. A. Rubio (Max Planck Institute), Dr. Y.-F. Zhao (University of Washington) for helpful discussions.

**Author Contribution**

Z.W. and M.F.C. initiated and conceived the research. R.Z. performed on-surface growth and STM characterization with support from Z.W., Y.Y., Y.Q. and M.F.C. Z.W., R.Z. H.P., F.W. and M.F.C. analyzed the STM data. H.P., Y.Q., D.B., Y.Y., A.H., F. W., R.Z. and A.S. carried out the device fabrication. W.T. and S.G.L. developed the theory and performed DFT calculations, as well as assisted with data interpretation. K.W. and T.T. grew hBN crystals. F.W., S.G.L. and M.F.C. acquired the funding and supervised the project. All authors contributed to scientific discussions and writing of the manuscript.

**Competing Interest**

The authors declare no competing interests.

**Data, Code and Materials Availability**

All data reported in the manuscript is provided as part of source data files. The DFT code can be downloaded from the Quantum Espresso website[43,44] and the pseudopotentials can be downloaded from the pseudo-dojo website[45,46]. For this study we used version 7.2 for the DFT–PBE calculations. The Wannier90 code can be downloaded from the Wannier90 website[47]. The tight-binding simulation code can be downloaded from the PythTB website (we used version 1.8.0 for extracting tight-binding parameters from DFT results)[48].

## Figures

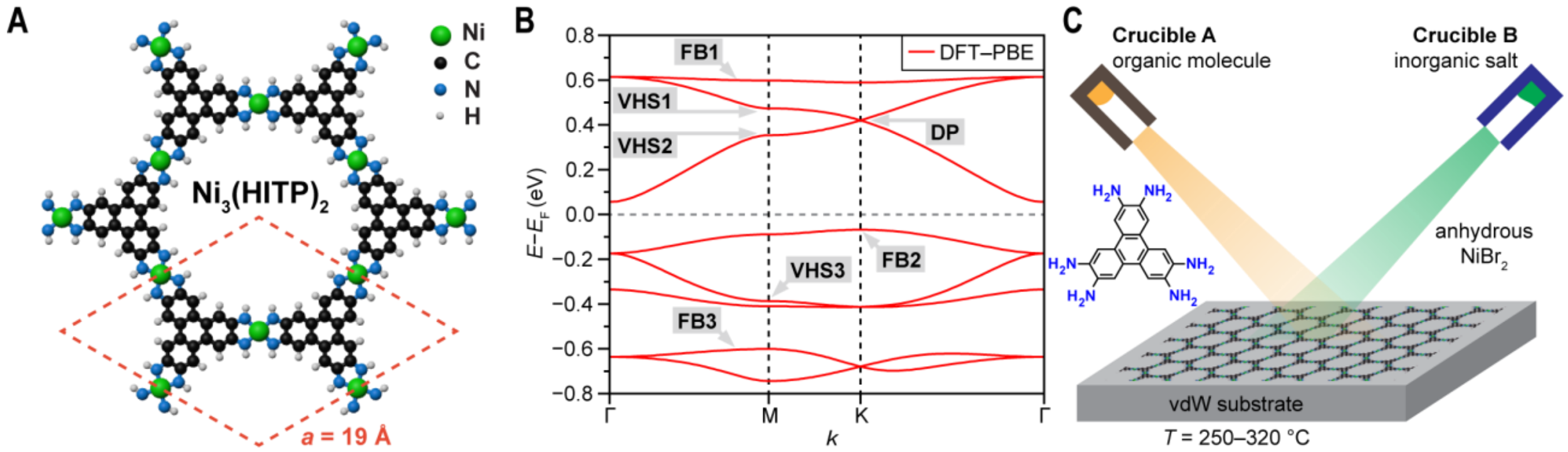

**Fig. 1. $Ni_3(HITP)_2$ structure and metal-salt MBE growth.** (A) Sketch of the $Ni_3(HITP)_2$ kagome metal–organic framework (MOF) chemical structure. Square-planar Ni centers are linked by HITP ligands to form a network of corner-sharing triangles. (B) DFT–PBE band structure of a freestanding $Ni_3(HITP)_2$ monolayer shows kagome-derived flat bands (FBs) and van Hove singularities (VHSs). (C) Schematic of metal-salt MBE growth of $Ni_3(HITP)_2$ on a van der Waals substrate.

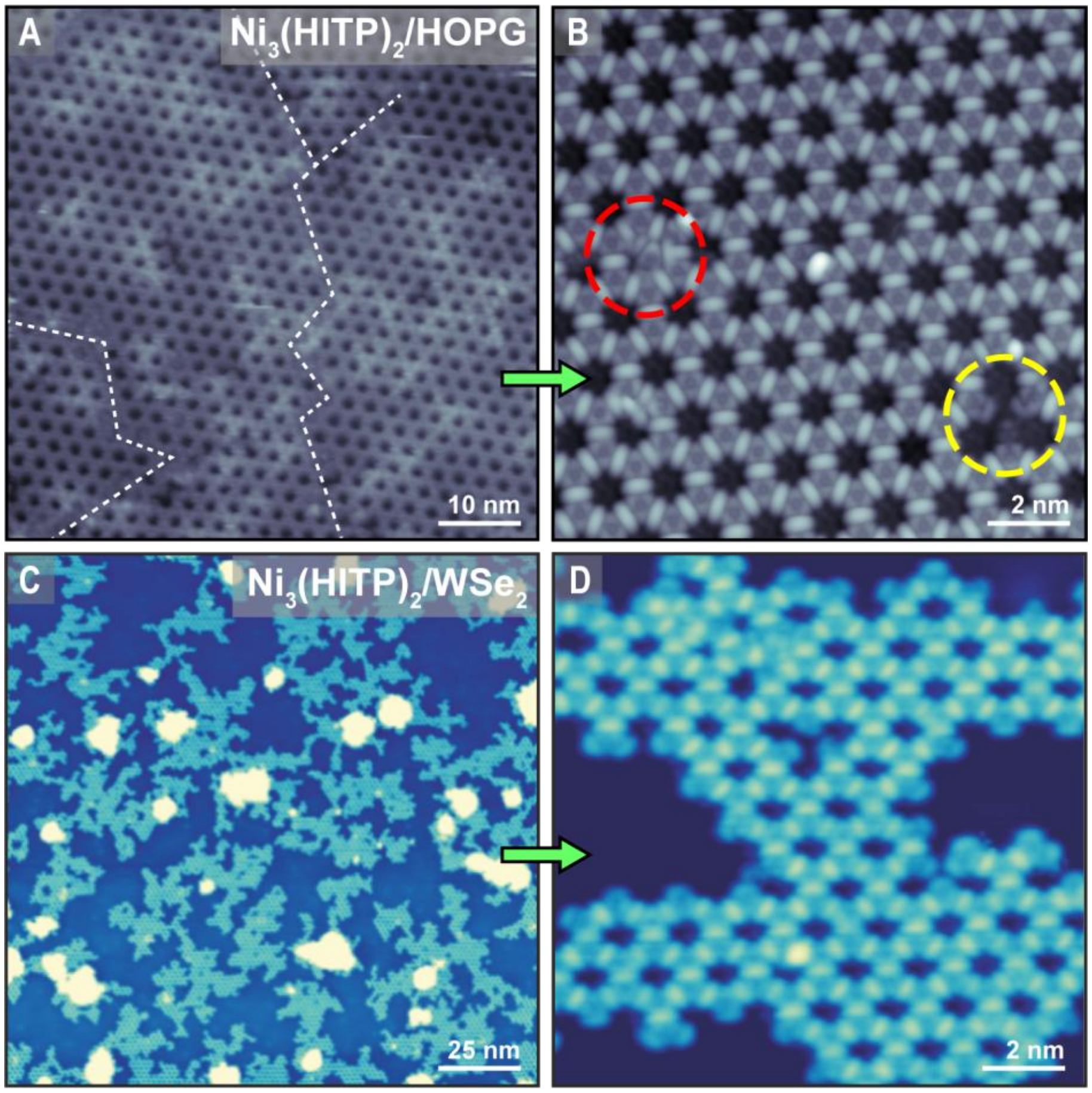

**Fig. 2. Metal-salt MBE growth of $Ni_3(HITP)_2$ on different van der Waals substrates.** (A) Large-area STM topographic image of an as-grown $Ni_3(HITP)_2$ monolayer on HOPG with grain boundaries indicated by white dashed lines (sample bias $V_s$ = +500 mV, tunnel current $I_t$ = 30 pA, $T$ = 4.7 K). (B) Close-up STM topographic image of the same $Ni_3(HITP)_2$ monolayer on HOPG with red and yellow dashed circles indicating interstitial molecule and molecular vacancy defects ($V_s$ = –1.0 V, $I_t$ = 50 pA, $T$ = 4.7 K). (C) Large scale STM image of $Ni_3(HITP)_2$ MOF grown on cleaved bulk $WSe_2$ crystal showing dendritic-like growth ($V_s$ = –1.5 V, $I_t$ = 15 pA, $T$ = 77 K). (D) Close-up STM topographic image of the same $Ni_3(HITP)_2$ monolayer on $WSe_2$ ($V_s$ = –1.2 V, $I_t$ = 25 pA, $T$ = 77 K).

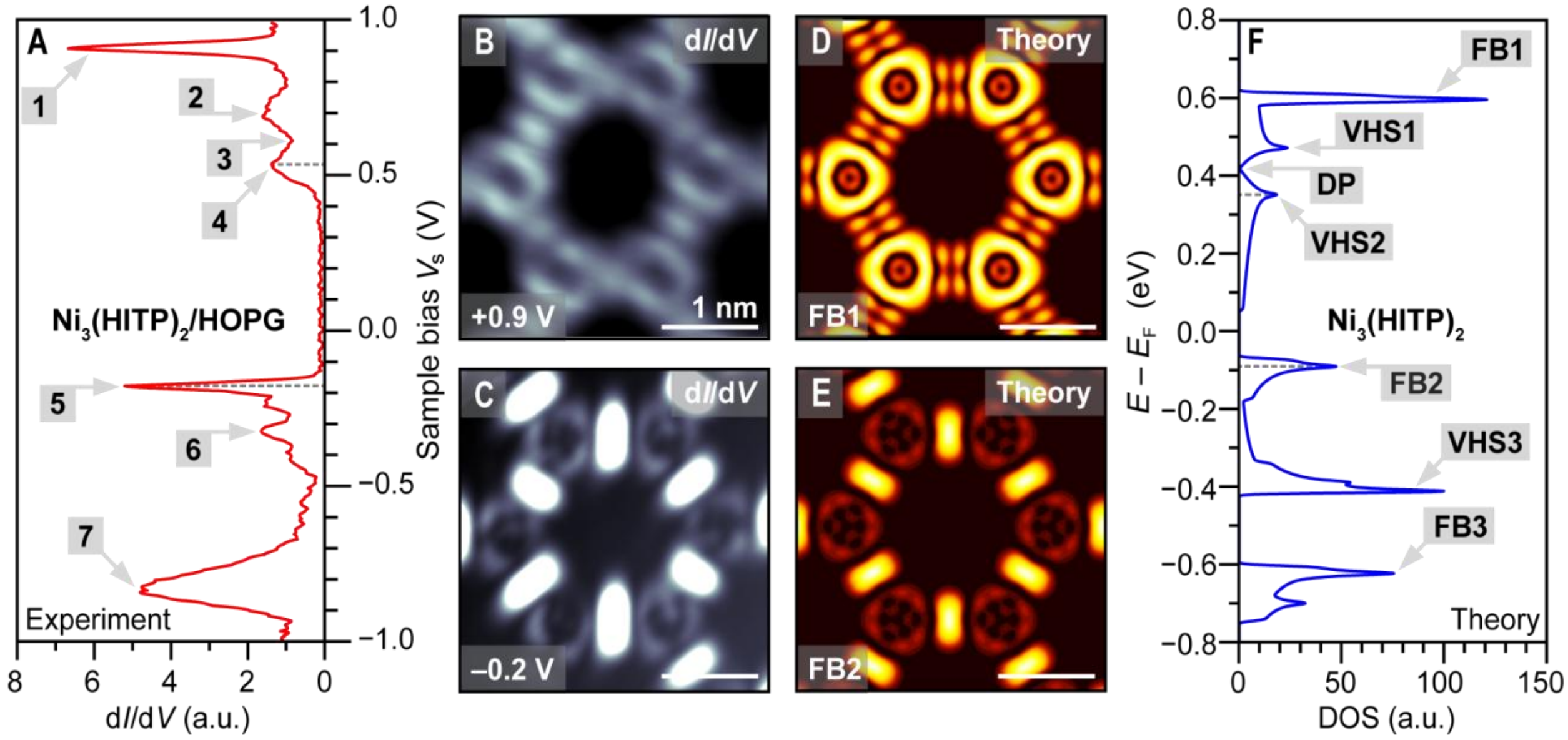

**Fig. 3. Spectroscopic characterization $Ni_3(HITP)_2$ electronic structure.** (A) STM d*I*/d*V* point spectroscopy of $Ni_3(HITP)_2$ on HOPG (spectroscopy: $V_{ac}$ = 5 mV, $f$ = 571 Hz; setpoint: $V_s$ = 1 V, $I_t$ = 100 pA, $T$ = 4.7 K). (B,C) Constant-height d*I*/d*V* maps recorded at (B) $V_s$ = +900 mV and (C) $V_s$ = –200 mV ($V_{ac}$ = 30 mV, $f$ = 571 Hz, $T$ = 4.7 K). (D,E) Calculated DFT–PBE LDOS maps above freestanding $Ni_3(HITP)_2$ at energies corresponding to (D) FB1 and (E) FB2 (sampled at a height 3.5 Å above the atomic plane of the $Ni_3(HITP)_2$). (F) Calculated DFT–PBE DOS for freestanding $Ni_3(HITP)_2$ (spectrum broadened by 13 meV Gaussian).

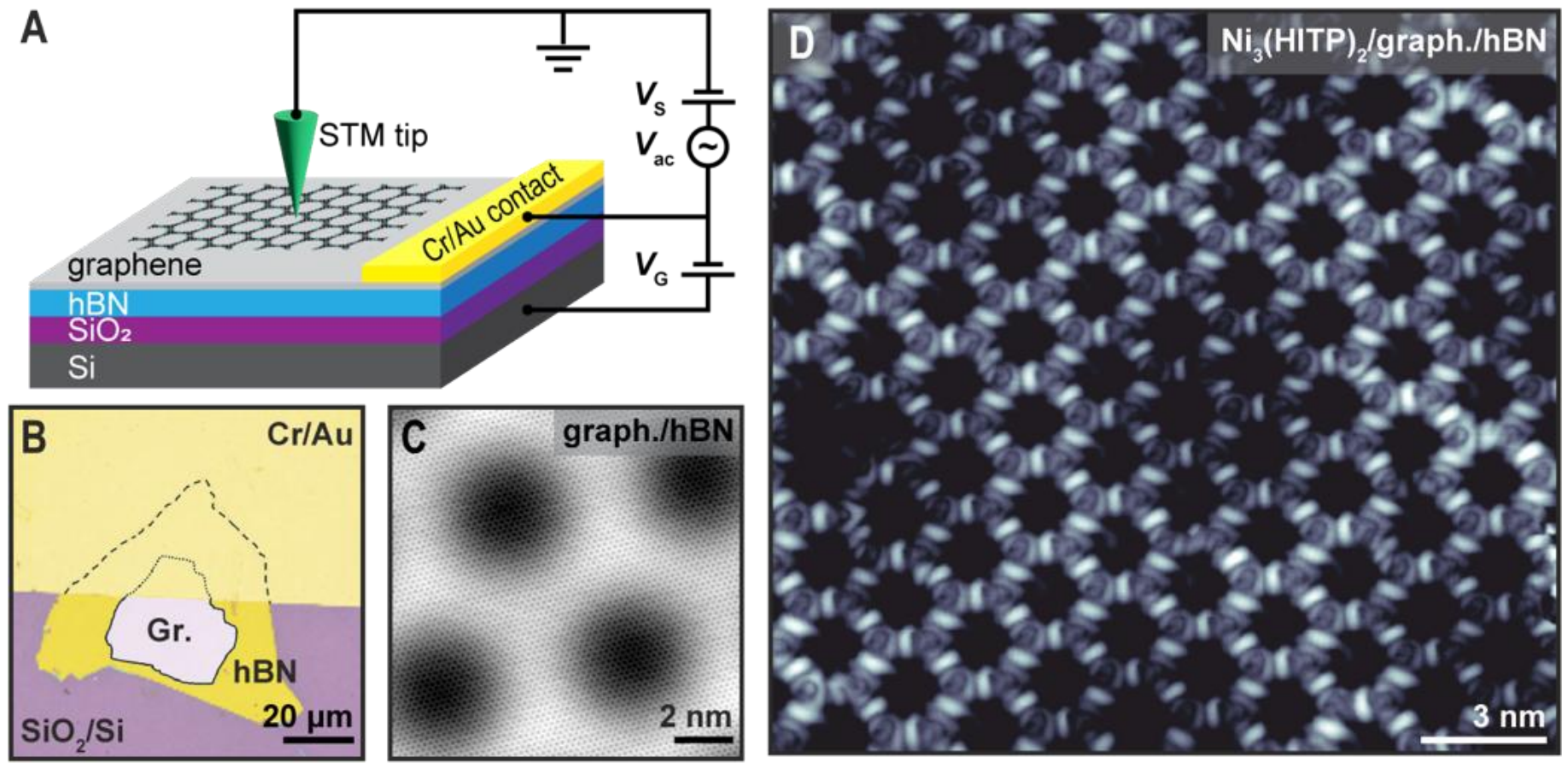

**Fig. 4. Metal-salt MBE growth of $Ni_3(HITP)_2$ on monolayer graphene/hBN devices.** (A) Sketch of experimental setup for growth of single-layer $Ni_3(HITP)_2$ on monolayer graphene/hBN device integrated with STM. A sample bias $V_s$ with superimposed wiggle voltage $V_{ac}$ is applied between the Cr/Au contact and the STM tip, and a back-gate voltage $V_G$ is applied between the doped Si wafer and Cr/Au contact. (B) Top-view optical micrograph of monolayer graphene/hBN with Cr/Au contact also visible. (C) STM topographic image of the monolayer graphene/hBN device *before* $Ni_3(HITP)_2$ growth shows atomic resolution as well as a much larger moiré pattern between graphene and hBN ($V_s$ = –1V, $I_t$ = 50 pA, $T$ = 4.7 K). (D) STM topographic image of the same device shown in (B, C) *after* growth of single-layer $Ni_3(HITP)_2$ ($V_s$ = –1V, $I_t$ = 20 pA, $T$ = 4.7 K). The large-scale (l ~ 9nm) modulation originates from the graphene–hBN moiré superlattice.

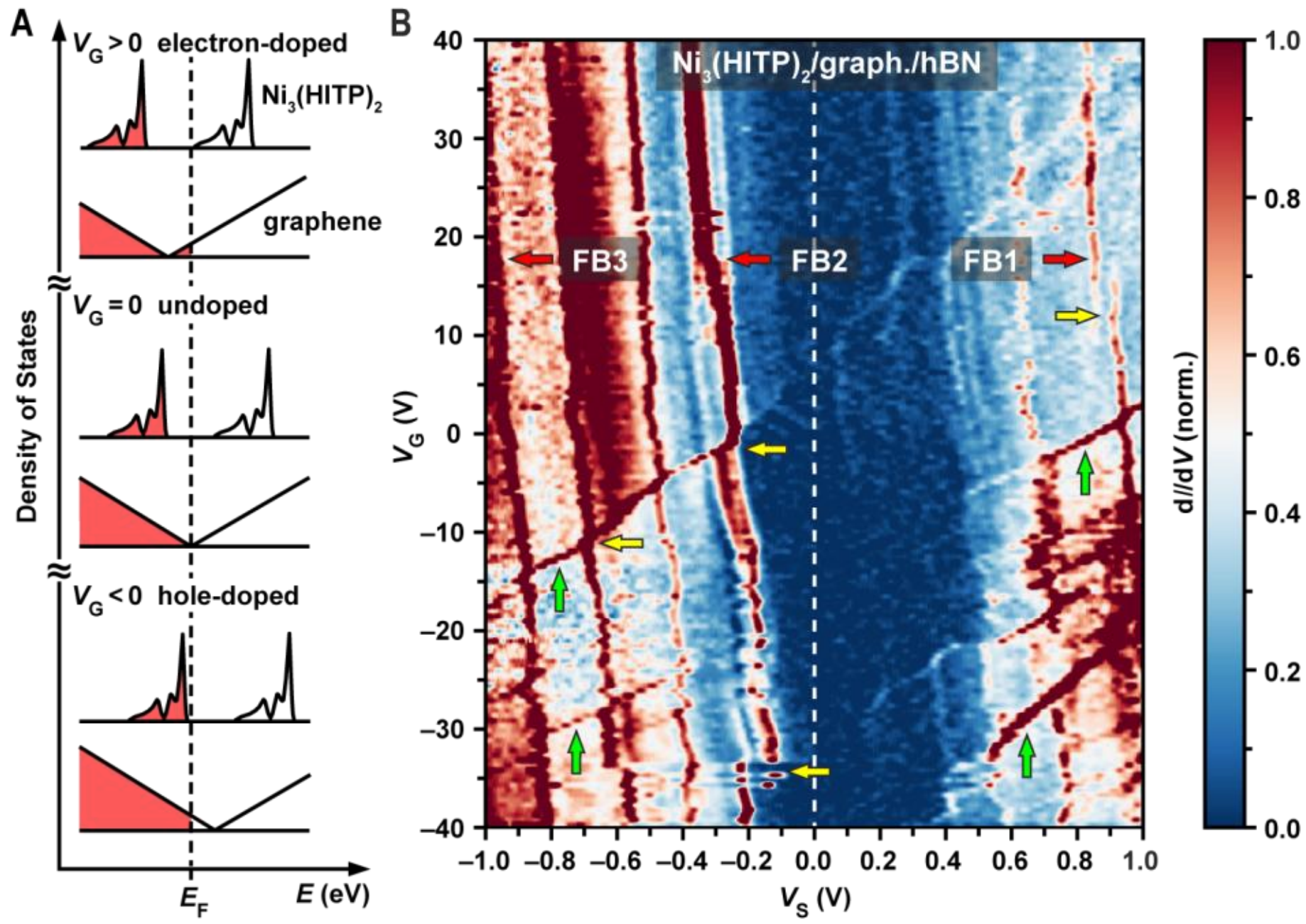

**Fig. 5. Gate-tunable kagome electronic structure: dependence of d*I*/d*V* on $V_s$ and $V_G$.** (A) Cartoon shows expected change in electronic structure induced by electrostatic gate voltage ($V_G$) assuming rigid band shift. $V_G > 0$ corresponds to electron-doping (DOS shifts to the left), while $V_G < 0$ corresponds to hole-doping (DOS shifts to the right). (B) Experimental density plot shows $V_G$ and $V_s$ dependence of d*I*/d*V* for single-layer $Ni_3(HITP)_2$ on a monolayer graphene/hBN FET device ($V_{ac} = 7$ mV, $f = 571$ Hz, $T = 4.7$ K). Band structure features (red arrows) and charging features (green and yellow arrows) are marked. The $Ni_3(HITP)_2$ energy gap (dark blue) is clearly visible in the center region of the plot.